# Automatic Parking in Smart Cities

Arezou Abyaneh
Electrical Engineering department
Qatar University
Doha 2713, Qatar
az1303286@qu.edu.qa

Vanessa Fakhoury
Department of Electrical and Information Technology
Lund University, Sweden
va5173fa-s@student.lu.se

Nizar Zorba
Electrical Engineering department
Qatar University
Doha 2713, Qatar
nizarz@qu.edu.qa

*Abstract*—The objective behind this project is to maximize the efficiency of land space, to decrease the driver stress and frustration, along with a considerable reduction in air pollution. Our contribution is in the form of an automatic parking system that is controlled by cellular phones. The structure is a hexagon shape that uses conveyor belts, to transport the vehicles from the entrance into the parking spaces over an elevating platform. The entrance gate includes length-measuring sensors to determine whether the approaching vehicle is eligible to enter. Our system is controlled through a microcontroller, using cellular communications to connect to the customer. The project can be applied to different locations and is capable of capacity extensions.

## I. INTRODUCTION

In the modern world, with a rapidly growing population, the parking of vehicles became a major problem faced every day. This issue results in the growth of traffic on roads, air pollution and driver frustration. Vehicles are parked along streets, causing congestions due to the fact there are insufficient vacant spots, or there is a very low efficiency in terms of the usage of space. An engineering solution is needed to avoid all these negative factors. The purpose is, increasing the efficiency of space used, as well as reduction of air pollution, frustration and stress which all drivers experience regularly while searching for parking spots.

As the era of automation and computerization is spreading at a high rate [1], the solution would take benefit of wireless communication, embedded systems, control, computer and electronic technology to implement an intelligent parking management system. The aim of automatic parking is to improve user experience and save precious time, resulting in the benefits mentioned above.

The method of automatic parking consists of placement and retrieval of cars by a computerized and programmed system. The customer will have the luxury of retrieving his/her vehicle through a regular cellular phone by sending a confirmation SMS to the system, which will automatically start the retrieval process and return a message containing billing information to the client, similar to the SMS sent by utility providers. The parking structure can be customized to fit specific locations, such as hospitals, malls, airports and even residential or commercial buildings, unrestricted to the size of the available land space or activity around it.

Despite the fact that automatic parking structures have been around for a while, they are not as common as they should be. The famous parking tower for Volkswagen in Wolfsburg, Germany includes an automatic parking, but the ability to retrieve is not available, as it is not designed for customers.

Previous students' works have tackled such problem, with different approaches and results. The work in [2] shows a vertical multilevel parking with three floors and two spaces on each floor. The device that will lift the car can move horizontally with help of a lead screw connected to a motor. Another motor to ramp up and down assures vertical movement. There is a cart attached to the device for placing the car into the spot and recovering it. Another group developed a semi-automated system [3] that will only elevate the car to different levels, and the driver needs to park the car into the free spots within that level. As the car approaches the entrance, a sensor sends a signal to the control system and the process starts. A Programmable logic controller helps the system to indicate the availability of the free spots with red and green lights. Another student project mentioned in [4] used AT89C52 microcontroller and fuzzy logic in Lab VIEW to achieve a secure, reliable and efficient car parking system. The building in this project has a circular shape with 3 floors and capacity for 24 cars. The system has two priorities for assigning the spots; size of the vehicle and urgency. According to this criterion, price will be decided. Within all previous designs in literature, our work is the first one that includes the car retrieval in advance through a cellular phone, adding a very important feature to make it fully automatic.

For our design, we have chosen a circular building structure with conveyor belts used for transportation of the vehicles; as well as the SMS service for the customers to request their cars in advance. We will use a DC motor and two pulleys for lifting and lowering the cars, while preventing friction. Other components of the system include a relay for switching among motors and an Inductive Proximity Switch, among others. A detailed discussion about every component and the reason of certain decisions will be tackled in detail along the rest of the paper.

## II. SYSTEM DESCRIPTION

The main components of the Automated Multilevel Parking system include an Arduino MEGA Microcontroller, which is the most important device, connecting everything together and controlling the entire system. The conveyor belts are rotating with the help of stepper motors, which are connected to the microcontroller through a Dual H-Bridge Motor. The system includes an elevation motor with a rotating platform, as well as entrance and exit bars, also connected to the Arduino through a Darlington Transistor Array. Additionally, the automatic multilevel parking includes car length measurement, done by distance measuring sensors and a 3G shield [5] with an antenna for the communication through SMS.

Figure 1 summarizes the complete process of parking and figure 2 shows the same for the car retrieval. The process of operation at entrance for parking and at the exit for retrieval is simple, as the car approaches; the sensors will check the car's length, as well as vacancy availability. If both conditions are met, the customer is requested to enter his phone number on a touch screen at entrance, the system proceeds by opening the entrance gate, initializing the parking timer, sending an SMS to the customer's entered number and starting the car transportation to its assigned parking spot through the conveyor belts.

When a retrieval SMS request is received from the customer, the system will stop the parking timer and start the retrieval process, which consists of lifting the platform and moving it to the designated space, recovering the vehicle and transporting it towards the exit area. One last SMS, containing billing information, is sent to the customer and the exit gate will be opened once the payment is done. As nowadays it is very unlikely for the GSM network system to fail, checking the delivery status of SMS would be implemented in a real-life scenario. Additionally, it is important to note that in a real-life scale development, further sensors would be necessary to detect any malfunction of the conveyor belt/s, which will trigger an alarm in the system, bringing the process to a halt until the problem is solved by service staff.

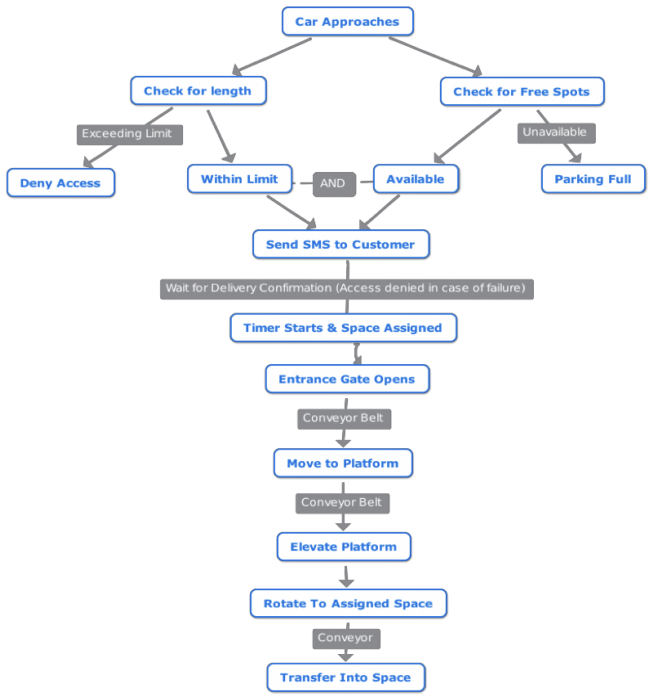

Fig 1. Parking process

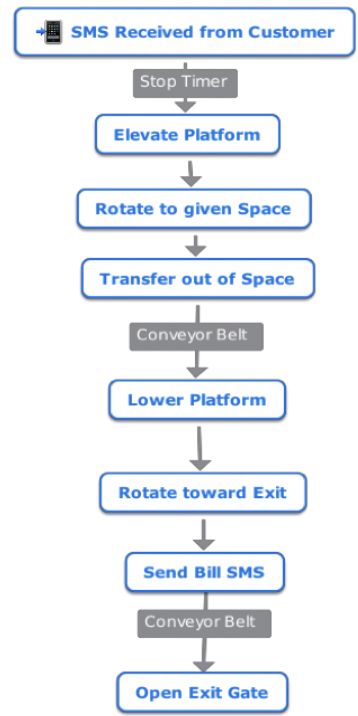

Fig 2. Retrieval process

## A. Constraints for Actual Implementations

An important step for any engineering project analysis is whether it tackles the design constraints imposed by the environment, and how it follows the available standards in the market, both locally and internationally.

General constraints of an automatic vertical parking include capacity, which is chosen in order to keep a safe distance between each vacancy on each level. Based on this safety distance, the system is restricted to the vehicles' dimensions. At the entrance of our automatic parking building, the length of the vehicle will be measured through sensors and not allowing an entry whenever this limitation is exceeded. Obviously, the limit can be changed to tackle alternative scenarios (e.g., trucks parking). The weight of the entering vehicle is also of great importance due to the elevating motors, as well as conveyor belt motors. Additionally, elevating and horizontal speed of the conveyor belts should be taken into consideration for the safety of the cars and to avoid collision with the structure itself. Conveyor belts were chosen for the transportation of the cars from the entrance onto the rising platform and into the spaces in order to cover a wide range of vehicle types, regardless of their dimensions, given they are within our restrictions.

Additionally, the parking and retrieval times are taken into consideration to avoid congestions and overcrowding around the building due to long waiting period. The parking/retrieval time depends on different factors such as the number of entrance and exit gates, number of elevators, parking spaces and floors. Based on current design parameters it was estimated to have a maximum of 2 minutes parking/retrieval time. Due to the fact that the structure is programmed by a controlled computer system, the starting and stopping points of the rotational platform and conveyor belt are fixed to fit the parking space's dimensions, which avoid the possibility of human error. Additionally, each slot will be equipped with parking buffers to avoid any damage.

At the drop-off location and once the vehicle is accepted, the customer enters his/her mobile phone number and a short SMS containing data, such as car park timing and information about billing are sent to the driver's phone. The GSM Communication standard was chosen, as it is sufficient for the used application to save printing out the parking card and bill. Other communication standards such as Wi-Fi are not applicable [6], due to the restriction in coverage as well as accessibility for users: all mobile phones support GSM but not necessary all of them allow internet connection.

III. HARDWARE SETUP

We will describe the hardware components which have been used in our project in the current section.

*A. Motors*

For conveyor belts motors, a detailed comparison was made between three types of motors that can be used for our application: Servo, stepper and DC motors. Summing up from comparison, stepper motors are the most suitable in this case, satisfying the two major property conditions such as the fact that there is no need of very high speed, while on other hand precise positioning is an essential requirement. Additionally, as there are nine belts in our design, the price of each motor was taken into consideration. Another benefit of stepper is the simplicity of its drive circuits, which are easy to control and set up. Accordingly, a hybrid stepping motor type 17HA7401-09 was chosen based on its high accuracy, low noise and smooth movement. Similarly, for driving this motor a dual full-bridge driver, L298N was used. The L298N is a high voltage, high current driver designed to accept transistor-transistor logic level and control stepper motors [6].

As of entrance and exit gates, stepper motors were programmed to move 90 degrees up and down to open and close the gate barrier, where a Portscape 26M048B unipolar coil was used. Some highlights about this motor are its excellent open loop control, easy coding and driving, and cost effective. For controlling and programming this motor an ULN2003a Darlington transistor arrays was used.

To elevate the vertical platform, a pk244-01a oriental motor was used. This motor has a 30 cm single shaft. It has a 1.8-degree step angle, which makes it very efficient, with high torque and gives it inertia capability. For driving the motor, one L298N stepper motor driver was used, and its functionality was controlled by microcontroller. For the rotation purpose, two 17HA7401-09 motors were used, with two equal size circular shape platform being attached to the elevation motor, and the two hybrid motors are placed on the lower platform, which will be fixed all the time. The same type of gear is attached to the motors shaft and the upper platform, making the upper platform to rotate.

*B. Sensors*

At the entrance of the parking building, the length of the present vehicle needs to be measured. We placed three sensors at the gate, and if all three sensors are activated, meaning an object is detected; the system will not allow the entry of the vehicle. For this purpose, we decided to use a distance sensor. Two sensors are available: Infrared and Ultrasonic. We selected the Infrared sensor due to its simplicity, as well as its much lower unit price. The only objective in this project for this prototype is detecting the presence of an object; therefore, neither a large range nor its functionality in sunlight is necessary. By changing the scenario, alternative sensors could be used.

*C. Communication Shield*

The automatic parking system includes an SMS notification service over GSM Communication. The customer will receive a short message, which includes the time of the drop off and simple retrieval instructions. When the driver is ready to retrieve the car, a reply message should be sent to the system to start the recovery process. Once the vehicle is ready for picked up, another message will be sent which will include the time of retrieval as well as the amount that is due to be paid. An Arduino 3G communication shield has been included due to its availability. Such shield is capable of a much larger number of services, such as the upload and download to web servers, as well as GPS tracking, that can be used in future work.

*D. Microcontroller*

In simple words Microcontrollers, are tiny computers on a single IC, which give us the ability to control the whole system with source coding and increase the project reliability. An initial comparison among four possible options of microcontroller was done to choose the best one for our parking system. As a result, Arduino MEGA was chosen to be the main microcontroller as it has an adequate number of digital pins for whole system. Its processor is sufficient for the commands what will be used. There is another microcontroller in system for 3G shield, for that application Arduino UNO is used and it is cascaded with the main Arduino MEGA.

*E. Power flow switch*

In this project there are several bipolar motors and each of them needs four Arduino Digital pins. For saving pins and being efficient during the process, as a maximum of two motors need to work at the same time, switches were utilized. Therefore, we used switches to turn on and off the power supply for each motor drive. To achieve this purpose, transistors and relays were used. One of the many applications of bipolar transistors is to control the power flow in and out of a circuit. This gives an advantage to control the supply going to each drive and corresponding motors. Despite some limitation, transistor switches have lower costs and are more reliable for this particular project.

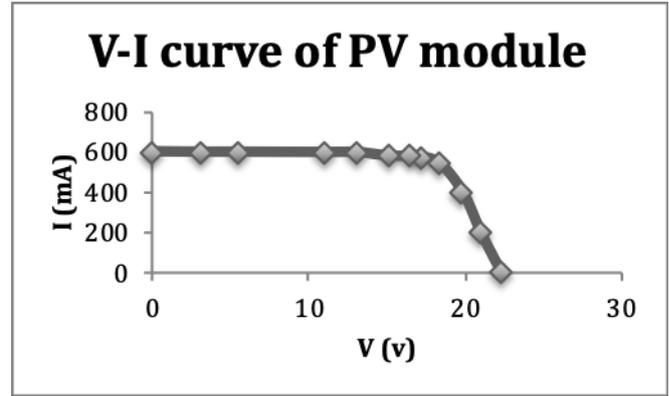

Fig 3. Label on the module

*F. Solar Panel*

Based on the environment of the Gulf region, it was decided to take advantage of the sun and convert it to electricity in this project. Including the addition of Solar panels onto the top of the building will increase energy efficiency; reduce the use of natural gas and $CO_2$ emission.

In this project, there are two parts that need a direct connection to power source: nine 12V stepper motors and two Arduino that need 7-12V to operate, the rest of the equipment like two gate motors and sensors are powered from the 5V output on the Arduino board itself [7].

A 30cmx30cm silicon solar PV module LS-10FX was provided as an alternative power source. Its specifications are as shown in Figure 3. First, we tested its performance by connecting it to CM 7371 LEDs, and the resultant V-I curve is plotted in Figure 4.
As it is shown in the Table 1, the specifications approximately match the values given on the module label in Figure 3.

For using solar panel in this project, a solar charge controller and sealed lead-acid battery is needed. For this purpose, a 3A, 12/24V solar charger controller was used to connect to all solar panel, battery and load units. We decided that solar energy only will be used for powering the motors and since there are nine motors in this project, and each has a rated power consumption of 10 watt per hour (W. h), it required a battery with a rated current consumption of higher than 7.5 Ampere per hour (Ah). But as in this project only 2 motors need to work at the same time, that reduces the amount of power consumption to around 2 Ah. Obviously, a backup source is needed as the solar energy is not available all the time, and a hybrid solution is rather more suitable for its implementation in realistic systems.

For expanding the idea to a real-life parking system there are some factors that should be considered. How much electricity does the system use per month, what percentage of the system should/could be powered by solar energy, how many peak sun hours does the building get during the day, how much roof space is available are among the main parameters. The lighting of the building, security system, fire system, air conditioning and many other systems are also additional sinks for the power.

Fig 4. VI curve of PV module

*G. Final Structure*

Figure 5 shows the complete model of the project, where all parts are connected together, which consist of conveyor belts, entrance and exit belt, elevating and rotating platform, entrance length sensor along with the solar panel, each one of them discussed in previous sections.

## IV. SOFTWARE SETUP

The Arduino Microcontroller allows the sensing and controlling of objects by programming through the open source IDE software, which is based on Java Programming. The code for this project consists of one main code and several functions. The functions include the reading of the sensors at the entrance gate, the motors' motion and the SMS sending/receiving. The functions feed information to the main code, allowing quicker processing. The main code on the other hand will check for available parking spaces using a simple matrix and assign, as well as to retrieve the data which has been obtained from the functions, in order to continue with data processing and decision making. Additional matrices are used as timers, to determine how long a vehicle has been parked.

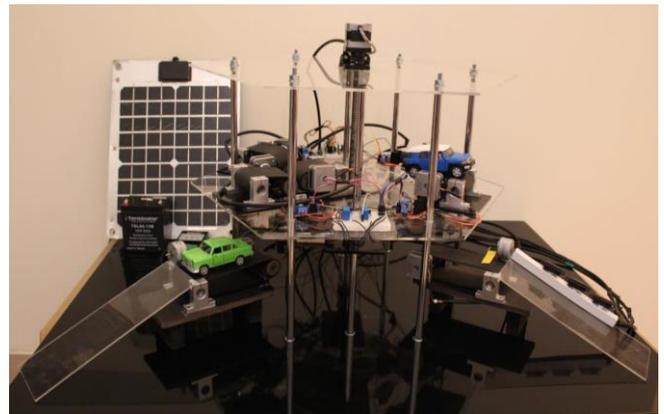

Fig 3. Full structure of the automatic parking

Table 1. Obtained value from Experiment

|  | Voltage (V) | Current (mA) |
|---|---|---|
| Open circuit | 22.31 | 0 |
| Maximum power | 18.36 | 540 |
| Short circuit | 0 | 601 |

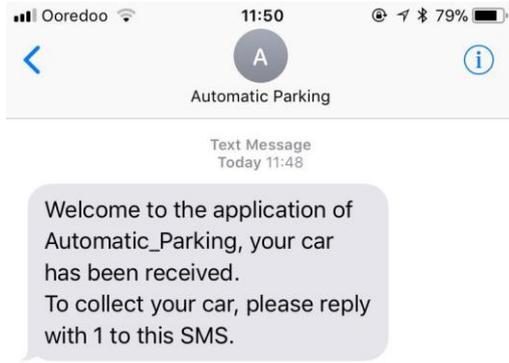

Fig 6. Received Customer SMS

Physical relay switches are used in the design, in order to simplify connections to the microcontroller, which are enabled by "digitally writing" HIGH's or LOW's of 5V or 0V respectively, to the given output pins. This step simplified the entire code and provides the ability to expand the setup for additional features.

*A. SMS*

In order to allow communication with the customer, an SMS will be sent at the entrance, which includes a welcoming message, as well as short instructions on how the vehicle can be retrieved as shown in Figure 6. The SMS is sent through a 3G shield, attached to an Arduino UNO, which is cascaded to the Arduino MEGA. The UNO was set up as a slave, while the MEGA is the master, only referring to the UNO when sending an SMS or waiting for a received one.

The Java based system uses AT commands such as AT+CREG=1, AT+CMGF=1, AT+CMGS=<number>, to connect to a home network, to set the SMS mode to text, or to send the SMS command in text mode respectively. The system is programmed to continuously check the inbox for a response from the customer, in order to initiate the retrieval process, depending on the customer's parking spot.

## V. CONCLUSIONS

Automatic Multilevel parking structures are of great advantages for busy cities, as they maximize ground space, lower the greenhouse emission and add to security, as well as simplicity of finding a vacant space. The project is designed in a way that allows expansion, by adding levels to the structure, as well as modifications to the software.

As additions to the current features of the system, the parking can be improved by extra options such as the development of a mobile application for the customer communication. The user could book a parking slot before arrival and if there are multiple automatic parking buildings around the city, the customer could check the nearest one with its availabilities. Finally, a website of the project is available through the URL VAParking.weebly.com, which includes videos of the processes, as well as images of all components and the structure.

## ACKNOWLEDGMENT

This work was made possible by NPRP grant NPRP 9-185-2-096 from the Qatar National Research Fund (a member of Qatar Foundation). The statements made herein are solely the responsibility of the authors.